# Field-free spin-orbit torque-induced switching of perpendicular magnetization at room temperature in WTe$_2$/ferromagnet heterostructures

Xiaomiao Yin[1], Lujun Wei[1, 2, *], Pai Liu[1], Jiajv Yang[1], Pengchao Zhang[1], JinCheng Peng[1], Fei Huang[1], Ruobai Liu[2], Jun Du[2,*], Yong Pu[1,*]

[1]*New Energy Technology Engineering Laboratory of Jiangsu Provence & School of Science, Nanjing University of Posts and Telecommunications (NUPT), Nanjing 210046, China*[1]

[2]*National Laboratory of Solid State Microstructures and Department of Physics, Nanjing University, Nanjing 210093, P. R. China*

**Abstract**

Spin-orbit torque (SOT) provides an efficient way to achieve charge-to-spin conversion and can switch perpendicular magnetization, which is essential for designing novel energy-efficient spintronic devices. An out-of-plane SOT could directly switch perpendicular magnetization. Encouragingly, field-free perpendicular magnetization switching of a two-dimensional (2D) material WTe$_2$/ferromagnet (FM) bilayer has been reported recently [1-4], but the working temperature (≤200 K) is below room temperature. Here, we report the field-free perpendicular magnetization switching carried out at room temperature on a WTe$_2$/Pt/Co/Pt multilayer film. Controlled experiments confirm that the field-free switching is caused by the in-plane antidamping SOT generated in the Pt/Co/Pt multilayer and the out-of-plane generated in the *a*-axis WTe$_2$ thin film. This work offers a potential method for using spintronic devices made of two-dimensional materials at room temperature.

---
[*] Corresponding Authors: wlj@njupt.edu.cn, jdu@nju.edu.cn, yongpu@njupt.edu.cn

**Introduction**

As a helpful method to induce magnetization switching[1,2,5,6], spin-orbit torque (SOT) has been the central topic of extensive discussion in manipulating electron spin due to its strength of low power consumption and high speed[7-10]. The traditional heavy metal (HM)/ferromagnet (FM) or topological insulators/FM heterostructures routinely transfer the in-plane current to the pure transverse spin current, which generated with spin polarization ($\sigma_y$) along the *y*-direction through the spin Hall effect[6,10-13]. In this case, the generated damping-like torque $\tau_y \sim m \times (m \times \sigma_y)$ only induced magnetization switching of in-plane direction. To deterministically switch a FM layer with perpendicular magnetic anisotropy (PMA), it is essential to apply an external magnetic field[6,10], an artificial asymmetric structure[14-17], an interlayer/exchange coupling[18], or a ferroelectric control[16] due to the help of additional symmetry breaking[11,19,20].

Instead, an out-of-plane SOT could directly switch perpendicular magnetization, which's spin polarization can generate an out-of-plane damping-like torque ($\tau_z \sim m \times (m \times \sigma_z)$). It can be generated by the interface engineering.[21,22] Recently, researchers have found some single materials generated an out-of-plane SOT, such as $Mn_2Au$[23], $Mn_3Ir$[24], $Mn_3Pt$[25] and $Mn_3Sn$[26] with low magnetic symmetry, CuPt[22] and $WTe_2$[27-31] with low atomic symmetry. Especially, two-dimensional (2D) transition metal dichalcogenides (TMDs) $WTe_2$ has rich spin-dependent properties[32-34]. The $T_d$-$WTe_2$ can be generated both in-plane spin current and out-of-plane spin current with unconventional spin orientation, and then induced non-traditional SOT along the out-of-plane, which can directly switch perpendicular magnetization switching without any magnetic field assistance[33-35]. It is worth noting that recent reports on the field-free perpendicular magnetization switching of a $WTe_2$/FM heterostructures are all realized at low temperatures (≤200K)[1-4]. Field-free magnetization switching of the $WTe_2$/FM heterostructures at room temperature has not been reported yet. In this work, field-free magnetization switching at room temperature is investigated using a $WTe_2$/Pt/Co/Pt multilayer film.

**Experimental Details**

**Device Fabrication.** First, a Pt(1.5)/Co(0.75)/Pt(1.5)/Cu(1) (FM) multilayer was deposited on cleaned Si/$SiO_2$ substrate by magnetron sputtering. The number inside parentheses indicates the thickness of each layer in the unit of nm, and all the thicknesses mentioned above are nominal. Then, the FM layer was patterned into a crossbar with a width of 5 μm by electron-beam lithography and ion etching. A thin $WTe_2$ flake of approximately rectangular shape was mechanically exfoliated on the crossbar of the FM layer (Fig. 1(a)). The exfoliation procedure was completed inside a nitrogen-filled glovebox with $H_2O$ and $O_2$ concentrations of 0.1 ppm. Finally, a h-BN flake is coated on the top of the $WTe_2$ flake to prevent oxidation of $WTe_2$. (Supplementary Fig. S1).

**Characterizations.** The thickness of the $WTe_2$ flakes was measured by the atomic force microscope (AFM, Dimension ICON). Angle-dependent Raman spectra (Renishaw inVia) using a micro-Raman spectrometer with a 633 nm laser excitation source. The in-plane and out-of-plane magnetic hysteresis (*M–H*) loops were measured by a commercial vibrating sample magnetometer (VSM, Microsense EV7). The magnetization switching and the anomalous Hall effect (AHE) loops were conducted with the home-made magnetic-electric system, where two Keithley source meters 6221 and 2182 and lock-in-amplifiers were employed. In current-induced magnetization switching measurements, for each data point, a pulsed d. c. electrical current with a duration of 50 μs was applied. After 5 s, the Hall resistance was recorded using a d. c.

excitation current (1 mA). We used the WTe$_2$ flake of the sample with two different thicknesses approximately 8.7 nm and 10.8 nm, which are denoted as device A and device B (Supplementary Fig. S9), respectively. All the tests were performed at room temperature.

**Results and discussions**

A schematic view of device A (Fig. 1 (a)) show the square WTe$_2$ was mechanically exfoliated on a Pt/Co/Pt/Cu/SiO$_2$/Si surface. The Pt(1.5nm)/Co(0.75nm)/Pt(1.5nm) multilayer film is chosen as the FM layer due to its excellent PMA and stability at room temperature. The thickness of the WTe$_2$ flake of device A is approximately 8.7 nm as measured by an AFM (Supplementary Fig. S1). The stacking structure of the device A is schematically displayed in Fig. 1 (b). The surface of WTe$_2$ is structurally mirrored symmetric only with respect to the *bc* plane, but not with respect to the *ac* plane. As a result, the 180° rotation of the system with respect to the *c*-axis of the is asymmetric[1].

The distinct on *a*-axis and *b*-axis in experiments generally reflect the structural variations for the WTe$_2$ flakes. The positions of characteristic peaks in the Raman spectra of the WTe$_2$ flake (Fig. 1 (c)) are consistent with those previously reported[36-38]. The angle-dependent polarized Raman spectrum of the device A's WTe$_2$ flake is shown in Fig. 1(d), and it is obtained by 180° counterclockwise rotation of laser polarization with respect to the long edge of the WTe$_2$ flake. Figure. 1(d) displays the intensity ratio of the $A_g^3$ and $A_g^1$ peak dependence on the polarization angle. The maximum and minimum values of the $I_{A_g^3}/I_{A_g^1}$ are observed at approximately 180° and 90°, respectively. We can therefore infer that the *a*-axis and *b*-axis of the WTe$_2$ flake (Fig. 1(a)) are the direction of the short and long sides, respectively.

The anomalous hall resistance (AHR) and current-driven magnetization switching are investigated. The out-of-plane AHR curve of the only FM (Pt/Co/Pt/Cu) sample (Fig. 2(a)) demonstrates the typical characteristic of PMA materials and $2R_{AH} = 5.1 \, \Omega$. Figures. 2(b) and 2(c) separately reveal the AHR of device A with the current along the *a*-axis and *b*-axis. We noticed that the $2R_{AH}$ of device A is slight smaller than that of the only FM sample due to the WTe$_2$ layer division of current characteristic.

Current-induced magnetization switching loops are measured at zero magnetic field for the FM sample (Fig. 2(d)), the device A with the current along the *a*-axis (Fig. 2(e)) and *b*-axis (Fig. 2(f)), respectively. It is noteworthy that the magnetization switching of the FM sample, the FM/WTe$_2$ heterostructures with the current along the *b*-axis cannot be switched at zero magnetic field, while the magnetization switching of the FM/WTe$_2$ heterostructures with the current along the *a*-axis can also be switched. At 50 Oe, the switching loops are obviously observed for this three cases (Fig. 2(g, h, i)). For the FM sample, the device A with the current along the *b*-axis, the magnetization can be switched with assistance of the external magnetic field, and their switching polarities change when the opposite magnetic field is applied (Supplementary Fig. S2). For the device A with the current along the *a*-axis, it is notable that the switching polarity, accompanied by a vanished switching ratio, does not appear at 0 Oe but at a small negative field of about −2.2 Oe (Fig. 2(d)), which show that the equal field is 2.2 Oe (Supplementary Fig. S2). Similar behavior as above are observed in the device B with the WTe$_2$ flake about 10.8 nm (Supplementary Fig. S10).

To further understand the switching behavior for the three cases, i.e. only FM sample, the FM/WTe$_2$ sample with the current along the *a*-axis and *b*-axis, respectively, controlled experiments are carried out at the zero field, as shown in Fig. 3. When the current is applied to the *a*-axis of the device A, whether the original magnetization state is set on spin-up state ($+M_z$) or spin-down state ($-M_z$) and the current is applied from zero to negative or from zero to positive, a counterclockwise switching loop can always be obtained (Fig. 3(b)), but a similar phenomenon cannot be observed for the FM sample (Fig. 3(a)) or the device A with the current along the *b*-axis (Fig. 3(c)). Even though a loop seems to be observed in Fig. 3 (c), it will disappear after repeated testing. On the contrary, when the current is applied the *a*-axis of device A, the loop is still oriented during repeated testing (Supplementary Fig. S4), which can infer that the $\sigma_z$ of WTe$_2$ plays an important role. Similar behavior as is also observed in the device B (Supplementary Fig. S11).

To more accurately estimate the contribution of WTe$_2$, we employ the first ($R_{1\omega}$) and second harmonic Hall resistance ($R_{2\omega}$) measurements. In the region, where the applied magnetic field $|H_x|$ is larger than $H_k$, $R_{2\omega}$ can be described by the following equation[9,40,41],

$$R_{2\omega} = \frac{R_{AH}}{2}\frac{H_{DL}}{|H_x|-H_k} + R_{offset} \qquad (1)$$

where $R_{AH}$ represent the AHR, $H_{DL}$ represent the SOT-induced effective field originating from damping-like torques, $H_k$ is magnetic anisotropy field and $R_{offset}$ is the offset signal. We obtain magnetic anisotropy field $H_k$ of about 732.37 Oe through numerically fitted the $R_\omega - H_x$ curve (Supplementary Fig. S5), which is consistent with the results of magnetic measurements of the FM film (Supplementary Fig. S8). Figure 4(a) presents the fitting curve of $R_{2\omega} - H_x$ through equation (1) at $H_x>$1 kOe or $H_x<$ -1 kOe, and then we can figure out the value of $H_{DL}$. The illustration of Fig. 4(a) shows $R_{2\omega}$ as a function of $H_x$ for the FM/WTe$_2$ sample with the current along the *a*-axis at an a. c. current amplitude of 3.5 mA and $\frac{\omega}{2\pi}$ =131.0 Hz. The current dependence of $H_{DL}$ is shown in Fig. 4(b) for the only FM sample. According to the linear fit, the spin-torque efficiency $\chi_{SOT}$ ($\chi_{SOT} = H_{DL}/J$) is calculated to be 0.198×10$^{-6}$ Oe/(Acm$^{-2}$), which is slightly greater than the value that has been reported for Pt/Co multilayers with a gradient of $0^9$. This may be due to the unequal thickness of the bottom and the top Pt films for the Pt/Co/Pt sample when the film is deposited by a magnetron sputtering system. When current is along the *a*-axis or *b*-axis (Fig. 4(c) and Supplementary Fig. S7), the $\chi_{SOT}$ of the WTe$_2$ flake is calculated to be $(-0.49\pm0.18)\times10^{-6}$ Oe/(Acm$^{-2}$) or $(-3.04\pm1.13)\times10^{-6}$ Oe/(Acm$^{-2}$) (Fig. 4(d) and Supplementary Note S2), which is consistent with what has been reported for the WTe$_2$/Ta/CoTb multilayer in the literature[42]. Note that, $\chi_{SOT}$ of the FM and FM/WTe$_2$ samples exhibit opposite signs. In addition, the $\chi_{SOT}$ of the device A is mainly contributed by Pt/Co/Pt multilayer (Fig. 4(d)), while is not the WTe$_2$ due to the current density ($J_{FM}=1.47\times10^7$ A/cm$^2$) at Pt/Co/Pt layer is approximately 14 times that ($J_{WTe_2}=0.105\times10^7$ A/cm$^2$) at the WTe$_2$ flake.

Afterwards, we calculated the damping-like effective Spin Hall angle ($\theta_{DL}$) using the formulas[43],

$$\theta_{DL} = \frac{2eM_s t H_{DL}}{\hbar j} = \frac{2eM_s t \chi_{SOT}}{\hbar} \qquad (2)$$

where *e* is the electron charge, $M_s$ is the saturation magnetization, *t* is the FM layer thickness and $\hbar$ is the reduced Planck constant. The $M_s$ value of the 3.75 nm-thick Pt/Co/Pt multilayer films is measured to be 158.4 emu/cm$^3$ (Supplementary Fig. S8). Calculated $\theta_{DL}$ of the FM is determined to be 0.0036, which is much less than the previously reported values in Pt/Co bilayer structure[10] due to the equivalence of the thickness the top Pt and the bottom Pt or interfacial Rashba effect. Similarly, when current is applied to *b*-axis or *a*-axis of the device A, calculated $\theta_{DL}$ of the WTe$_2$ layer is determined to be $(0.0089\pm0.0033)$ or $(0.055\pm0.020)$ (Fig. 4(d) and Supplementary Note 2 for detailed calculation process), respectively, which is comparable to the reported effective spin Hall angle at WTe$_2$[44]. The $\theta_{DL}$ of the WTe$_2$ layer is approximately 2.5 times and 15 times and that of the Pt/Co/Pt sample in current along the *b*-axis and the *a*-axis for the device A, respectively.

In a Pt/Co bilayers, the spin is polarized in *y* direction and the in-plane damping-like torque is presented as $\tau_y \sim m \times (m \times \sigma_y)$ (Fig. 5(a)), which can not only deterministically switch the magnetization of a magnet with PMA[12]. In this case, a field-free switch must be generated with the help of an external magnetic field. For the Pt/Co/Pt/WTe$_2$ structure in this paper, the Pt/Co/Pt multilayer can be generated in-plane anti-damping torque, since the WTe$_2$ flake can be generated *z*-polarized spin current when the current is applied along the low-symmetry *a*-axis[3,30]. The out-of-plane anti-damping torque exists in the form of $\tau_z \sim m \times (m \times \sigma_z)$) (Fig. 5(b)), eliminating the effect of external magnetic field and realizing the field-free deterministic magnetization switching.

**Conclusion**

In conclusion, we have experimentally investigated SOT-induced magnetization switching of the FM/WTe$_2$ heterostructures at room temperature. When current is along the a-axis, field-free magnetization switching can be obtained. On the contrary, the deterministic magnetization switching disappears at zero magnetic field for the FM sample, the devices with current along the *b*-axis. We obtained the $\theta_{DL}=(0.055 \pm 0.020)$ contributed by WTe$_2$ for the device A with the current along the *a*-axis, which is significantly higher than that of the Pt layer. Field-free switching is caused by the in-plane antidamping SOT generated in the Pt/Co/Pt multilayer and the out-of-plane generated in the *a*-axis WTe$_2$ thin film. Our findings pinpoint that WTe$_2$/FM opens great perspectives for the application of spintronic devices at room temperature.

**Acknowledgements**

This work was supported by the National Natural Science Foundation of China (Grant Nos. 52001169, 61874060, U1932159, 61911530220, 12104238), Natural

Science Foundation of Jiangsu Province (Grant No. 20KJB430010, 20KJB430017), NUPTSF (Grant Nos. NY219164, NY217118, NY219162), the open Project of the laboratory of Solid-state Microstructures of Nanjing University (Grant No. M33038), Foundation of Jiangsu Provincial Double-Innovation Doctor Program (grant no. CZ007SC20018), Jiangsu Specially-Appointed Professor program, Natural Science Foundation of Jiangsu Province (Grant No. BK20181388, 19KJA180007), Oversea Researcher Innovation Program of Nanjing, Innovation Project of Jiangsu Province (Grant Nos. KYCX20_0791, KYCX21_0697).

# Figure Captions

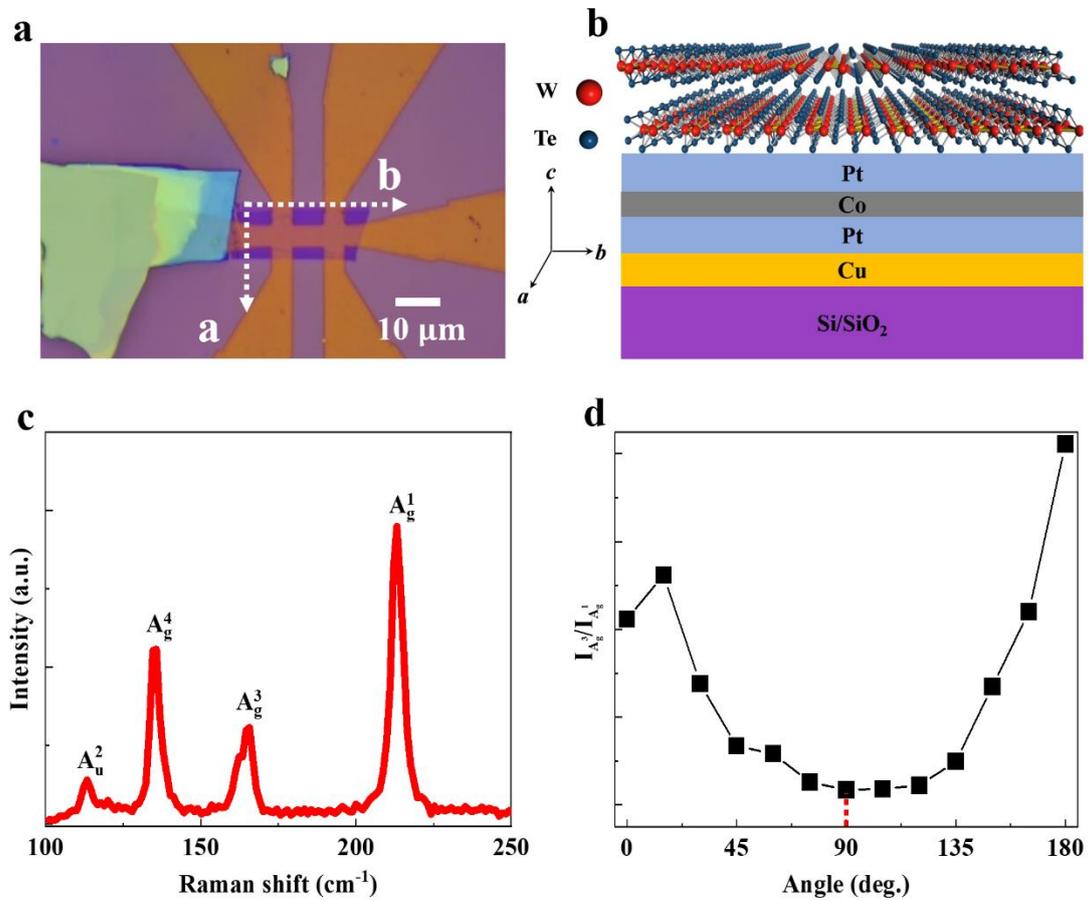

**Fig. 1 Structure, Raman spectra of the device A.** Optical picture (a) and stacking schematic (b) of the device A. Raman spectra (c) and angle-dependent polarized Raman spectral intensity of the WTe$_2$ flake of the device A.

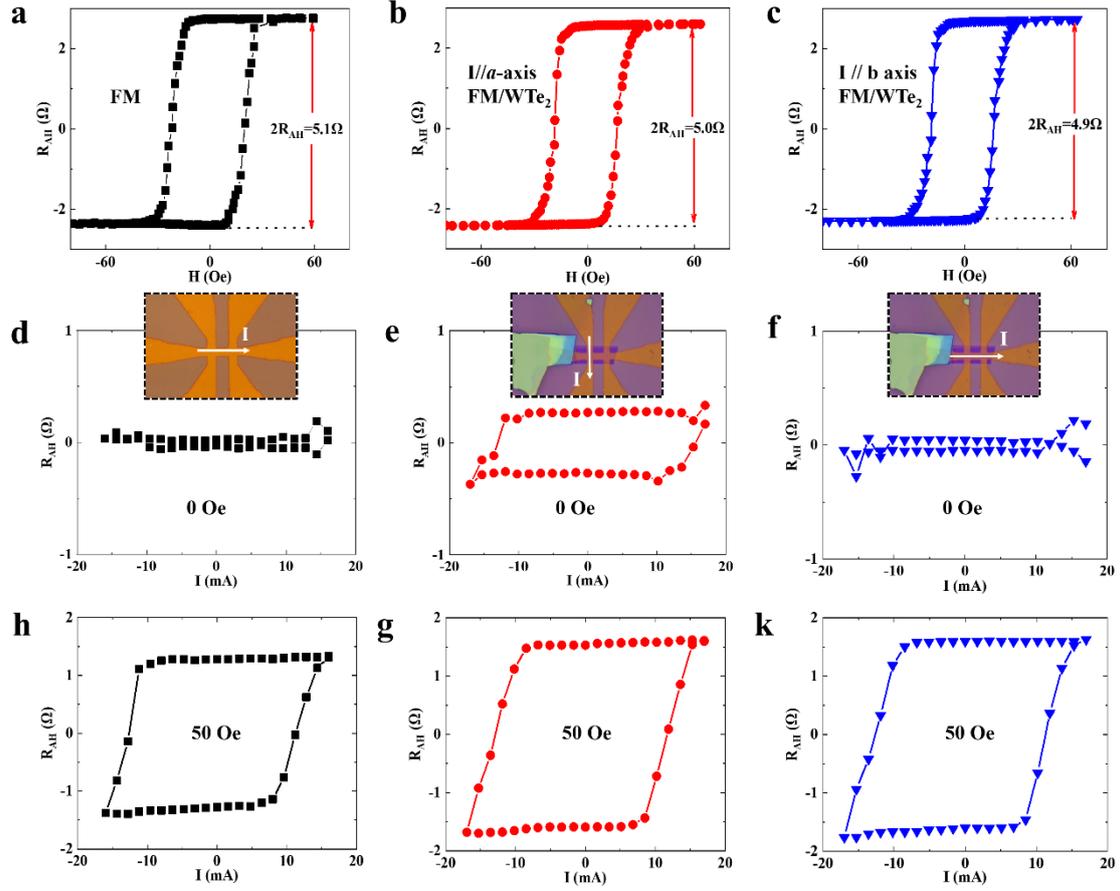

**Fig. 2 AHR and current-driven magnetization switching.** Out-of-plane AHR curves of the FM sample (a), the FM/WTe$_2$ sample (b, c) when current is along the *a*-axis and *b*-axis, respectively. $R_{AH}$ vs *I* curves measured at zero magnetic field of the FM sample (d), the FM/WTe$_2$ sample (e, f) when current is along the *a*-axis and *b*-axis, respectively. $R_{AH}$ vs *I* curves measured at 50 Oe of the FM sample (h), the FM/WTe$_2$ sample (g, k) when current is along the *a*-axis and *b*-axis, respectively. The insets of Figs. (d), (e) and (f) are shown the sample are measured by using the current direction along the arrows.

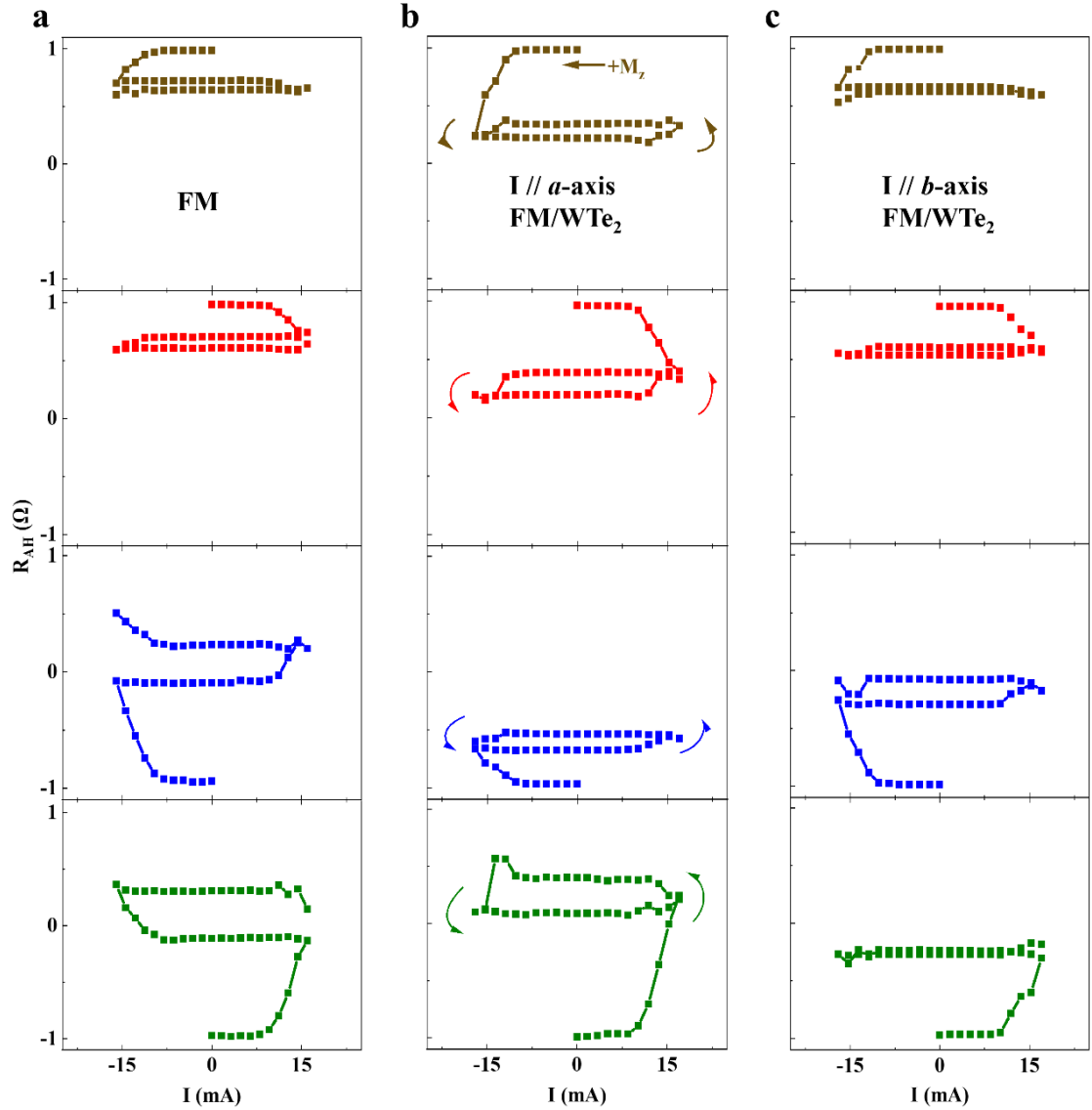

**Fig. 3 Current-induced switching for different initial states at zero magnetic field.** For the initial state of $+M_z$ or $+M_z$, and current from 0 to negative or positive measurements. Current-induced switching for the FM sample (a), device A with current along the *a*-axis (b) and *b*-axis (c), respectively. When current flows along the *a*-axis of device A, a counterclockwise loop is always obtained.

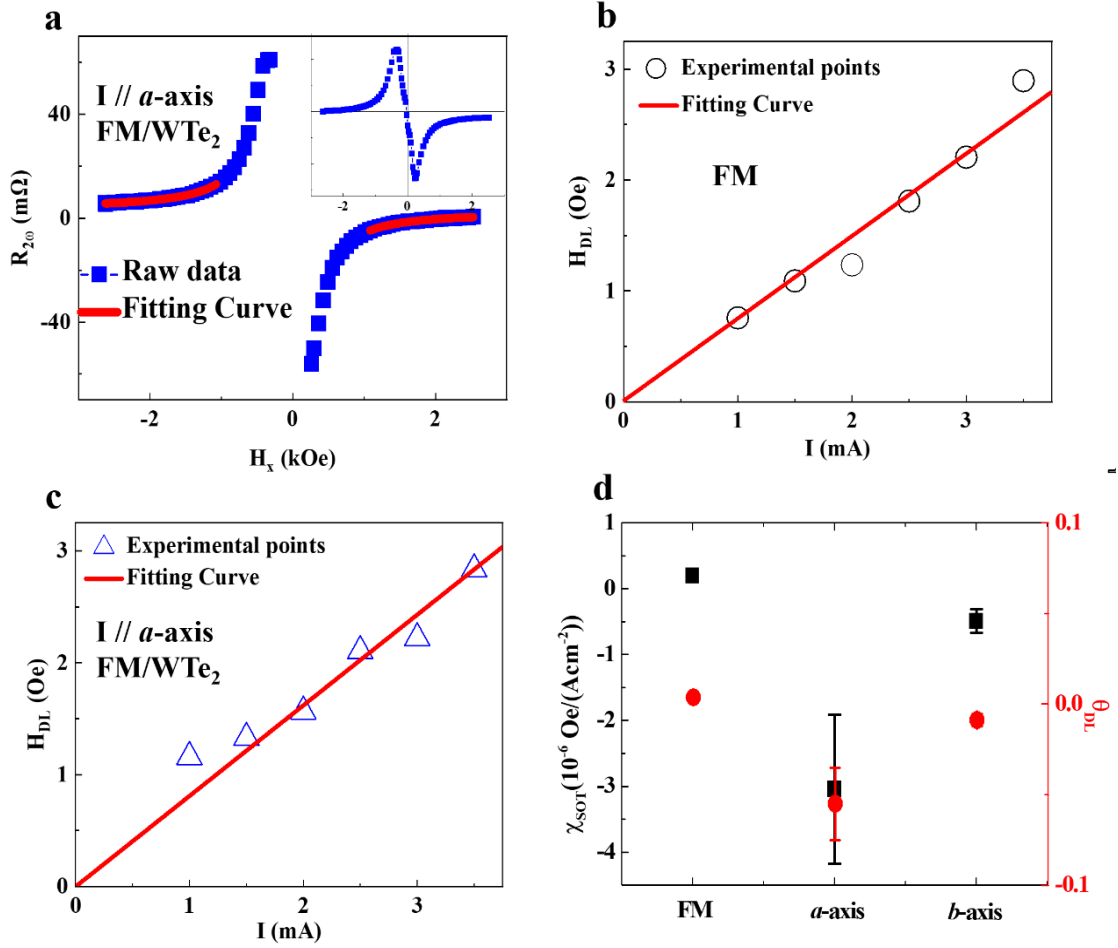

**Fig. 4 SOT efficiency characterization.** (a) $R_{2\omega}$ as a function of $H_x$ for the FM/WTe$_2$ sample with the current along the *a*-axis. (b) $H_{DL}$ as a function of current (*I*) for the FM sample (b) and the FM/WTe$_2$ sample (c) with the current along the *a*-axis. The slope of the red linear fitting curve represents $\frac{\chi_{SOT}}{S}$=0.835 Oe/mA for the FM sample, where *S* is the cross-sectional area of the FM layer Hall bar. (d) $\theta_{DL}$ and $\chi_{SOT}$ values for the FM sample, the FM/WTe$_2$ sample with the current along the *a*-axis and b-axis, respectively.

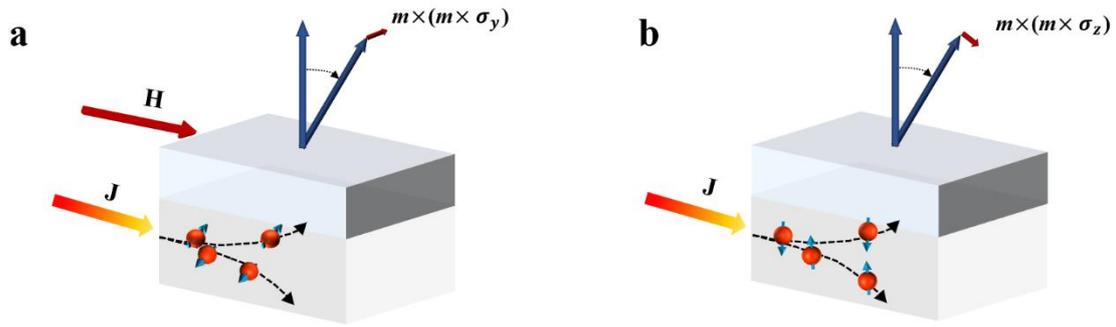

**Fig. 5 Mechanism of Switching by in-plane and out-of-plane damping-like SOTs.** (a) A schematic of a conventional HM/FM bilayer SOT device. An IP charge current passes along *x* direction in the bottom HM layer, generates an OP spin current with *y*-polarized. This spin current exerts an IP damping-like torque $\tau_y \sim m \times (m \times \sigma_y)$ on the perpendicular magnetization in the top FM layer. In this case, a sizable external magnetic field is required for a deterministic switching. (b) A schematic of a WTe$_2$/FM bilayer SOT device supporting the OP anti-damping torque ($\tau_z \sim m \times (m \times \sigma_z)$), which can realize a field-free switching on the perpendicular magnetization in the FM layer, where the IP charge current generates an OP spin current with *z*-polarized spin.

**Supporting Information**

**S1 Optical picture, AFM of the device A.**

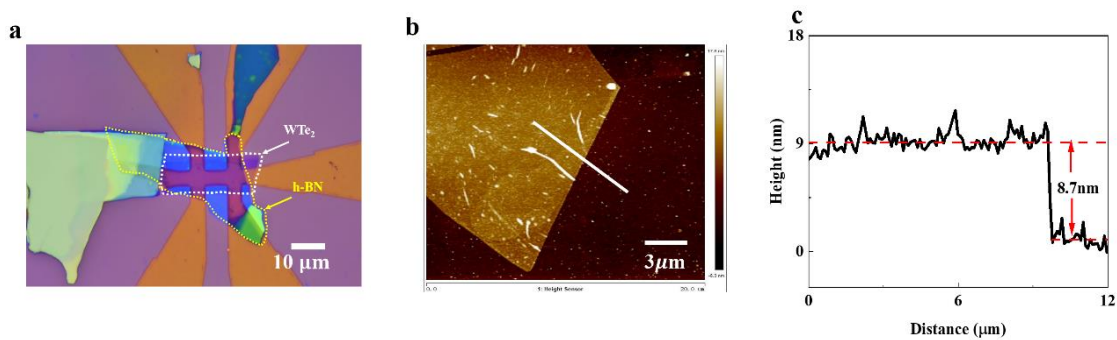

Figure S1. (a) Optical picture of the device A coated h-BN. The top h-BN prevents the WTe$_2$ layer from being oxidized. The dotted white box is the WTe$_2$ flake. The dotted yellow box is the h-BN flake. Scale bar is 10 $\mu$m. (b) An atomic force microscopy (AFM) image of the WTe$_2$ flake with similar thickness as WTe$_2$ flake used for fabrication of device A. (c) A line cut [white line in Fig. (b)] from the edge of the WTe$_2$ flake, presenting an average thickness of about 8.7 nm.

**S2 Current-induced switching for the FM sample, device A.**

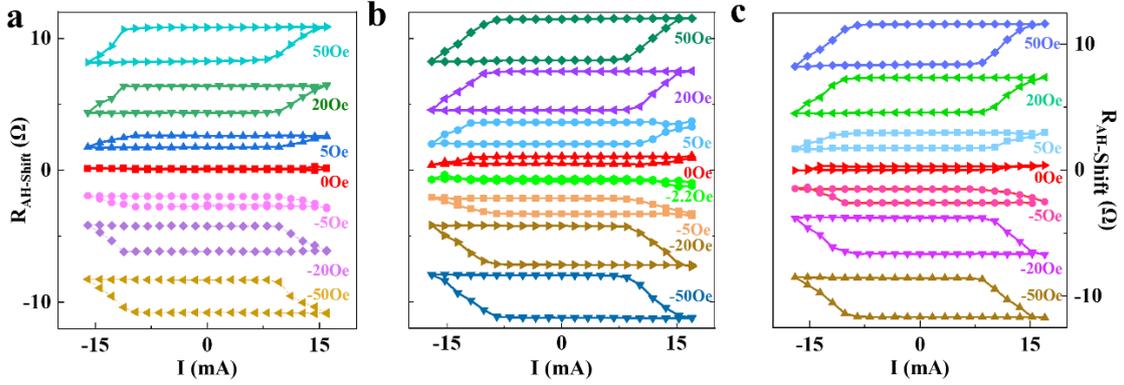

Figure S2. Current-induced switching loops at in-plane different fixed magnetic fields along the *x*-axis direction for the FM sample, device A with current along *a*-axis and *b*-axis, respectively.

The switching ratio increases as the magnitude of the field increase, with a maximum value of ~ 65% at 50 Oe for the device A with a current along the *a*-axis (Supplementary Fig. S3). Similar behavior is found for the only FM sample and the device A with the current along *b*-axis (Supplementary Fig. S3). The maximum value of switching ratio is 50% and 62% for the only FM layer and the device A with the current along the *b*-axis, respectively. The critical switching current density $J_C$ for the Pt/Co/Pt device is approximately $4.75 \times 10^7$ A/cm$^2$ with an in-plane magnetic field $H_x$=50 Oe, which $J_C$ is same as the previously reported values in Pt/Co multilayer structures. When the device A's current is flowing along its *a*-axis and *b*-axis for the device A, the critical switching current density is respectively donted as the $J_{Ca}$ and $J_{Cb}$, the $J_{Ca}$ and $J_{Cb}$ are $(4.92\pm0.006)\times10^7$ A/cm$^2$ and $(4.73\pm0.006)\times10^7$ A/cm$^2$ with an in-plane magnetic field $H_x$=50 Oe, respectively. (Supplementary Note S1 for detailed calculation process).

**S3 Switched ratio for the FM sample, device A.**

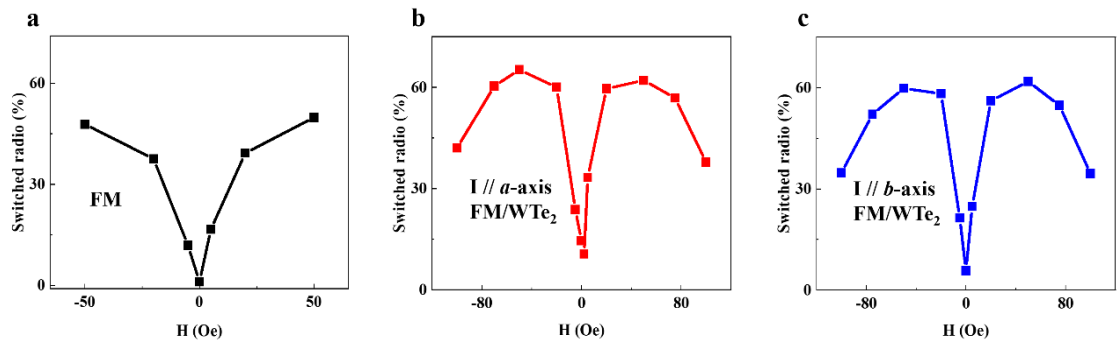

Figure S3. The switched ratio varies with in plane magnetic fields for the FM sample (a), the device A with current along the *a*-axis (b) and *b*-axis (c), respectively.

**S4. Current-induced switching at zero magnetic field before the magnetization to up for the device A.**

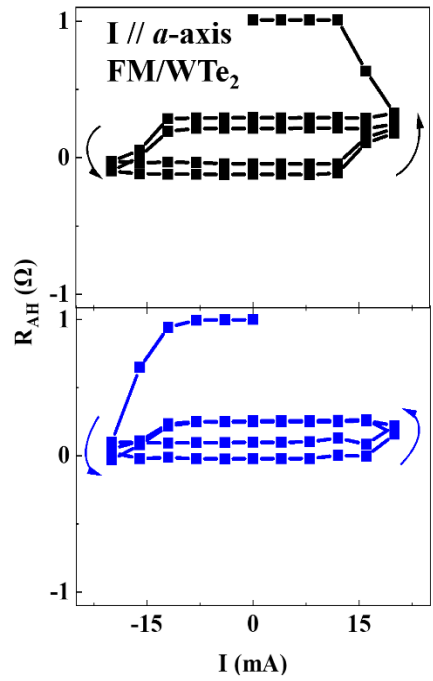

Figure S4. Under the condition of zero magnetic field, counterclockwise loops can always be obtained by applying the current from 0 to positive and from 0 to negative to the *a*-axis of device A in two consecutive measurements of the sample.

**S5 $R_{1\omega}$ and fitting of the FM sample.**

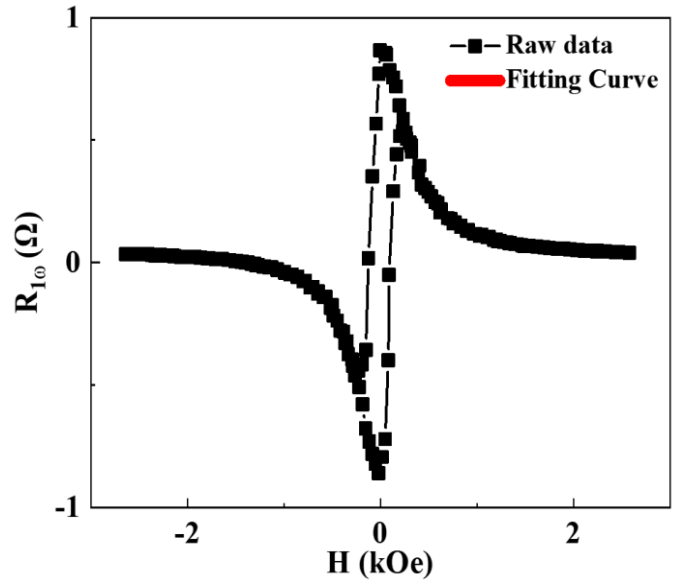

Figure S5. First harmonic Hall resistance ($R_{1\omega}$) as a function of $H_x$ and fitted curve for the FM sample.

Figure S5 displays the $R_{1\omega}$ of the only FM layer and device A at a small a. c. current amplitude of 1 mA and curve fitting of $H_k$. The fitting equation can be written as,

$$R/R_{AH}=\sin[\arccos(H_x/H_k)] \qquad (S1)$$

Where R is the AHR, $R_{AH}$ is the maximum of the AHR. Hx is the applied magnetic field. We obtain magnetic anisotropy field $H_k$ of about 732.37 Oe through numerically fitted the $R_{1\omega}$ curve.

## S6 $R_{2\omega}$ and fitting data of the FM sample and the device A.

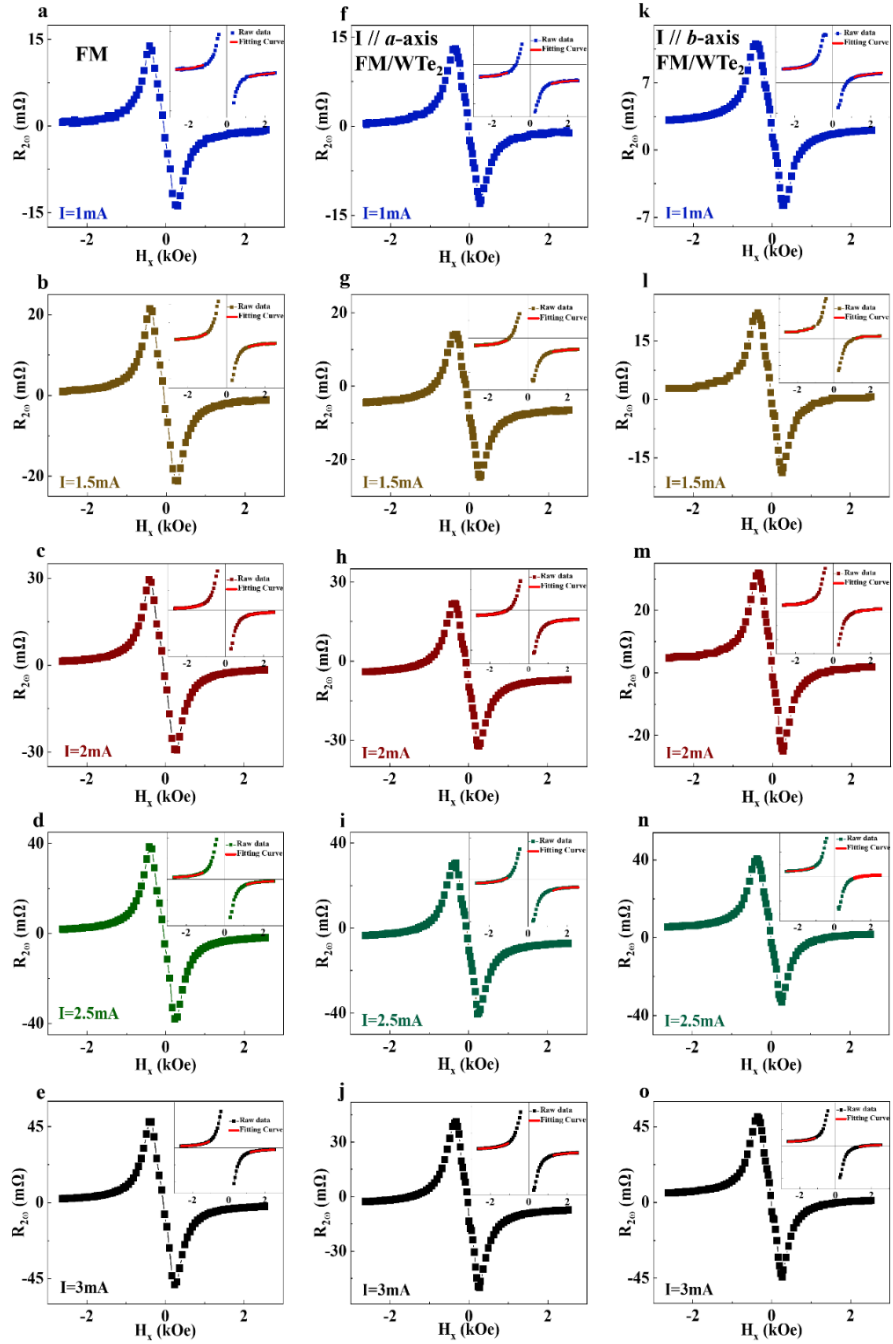

Figure S6. $R_{2\omega}$ as a function of $H_x$. (a) to (e) respectively show the variation of $R_{1\omega}$ with $H_x$ when the current is applied from 1mA to 3mA with only the FM sample. The inset shows the fitting curve at high field($H_x>$1 kOe or $H_x<$-1 kOe). (f) to (j) present the images of the second harmonic experiment when current from 1mA to 3mA is applied at the *a*-axis of device A. Using the same method to fit in the high field, as shown in the illusion. (k) to (o) show the images of the same test method when the current is applied to the *b*-axis of device A.

## S7 Fitting $H_{DL}$ of device A with current along the *b*-axis.

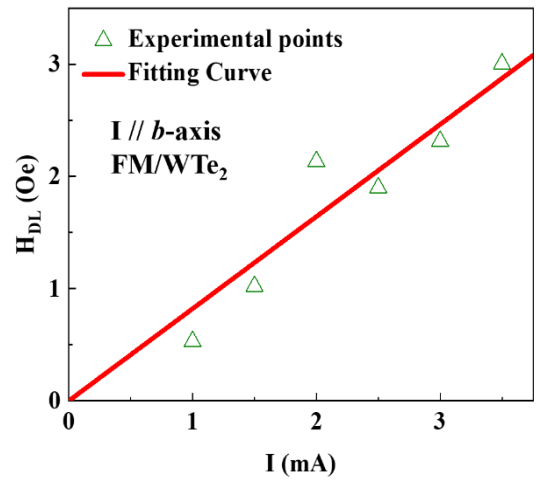

Figure S7. $H_{DL}$ as a function of current (*I*) of the device A for the I//*b*-axis. The slope of the red linear fitting curve is 0.8214 Oe/mA, where it includes $H_{DL}$ of the FM and WTe$_2$ layers.

**S8 *M-H* loops of the FM layer.**

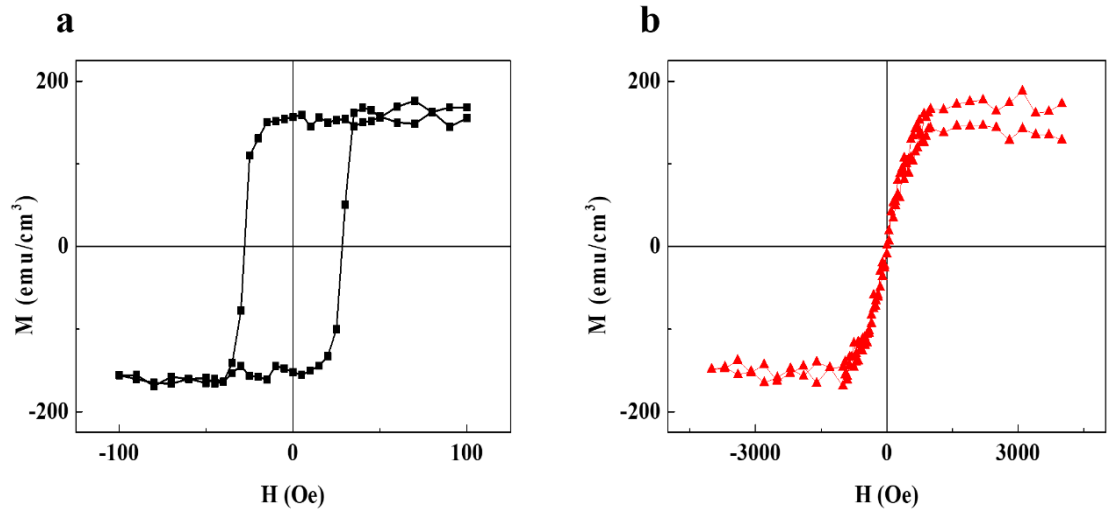

Figure S8. The out-of-plane (a) and in-plane (b) *M-H* loops for the Cu(1nm)/Pt(1.5nm)/Co(0.75nm)/Pt(1.5nm) multilayer. The saturation magnetization $M_s$= 158.4 emu/cm$^3$ can be obtained.

**S9 Optical picture, AFM of the device B.**

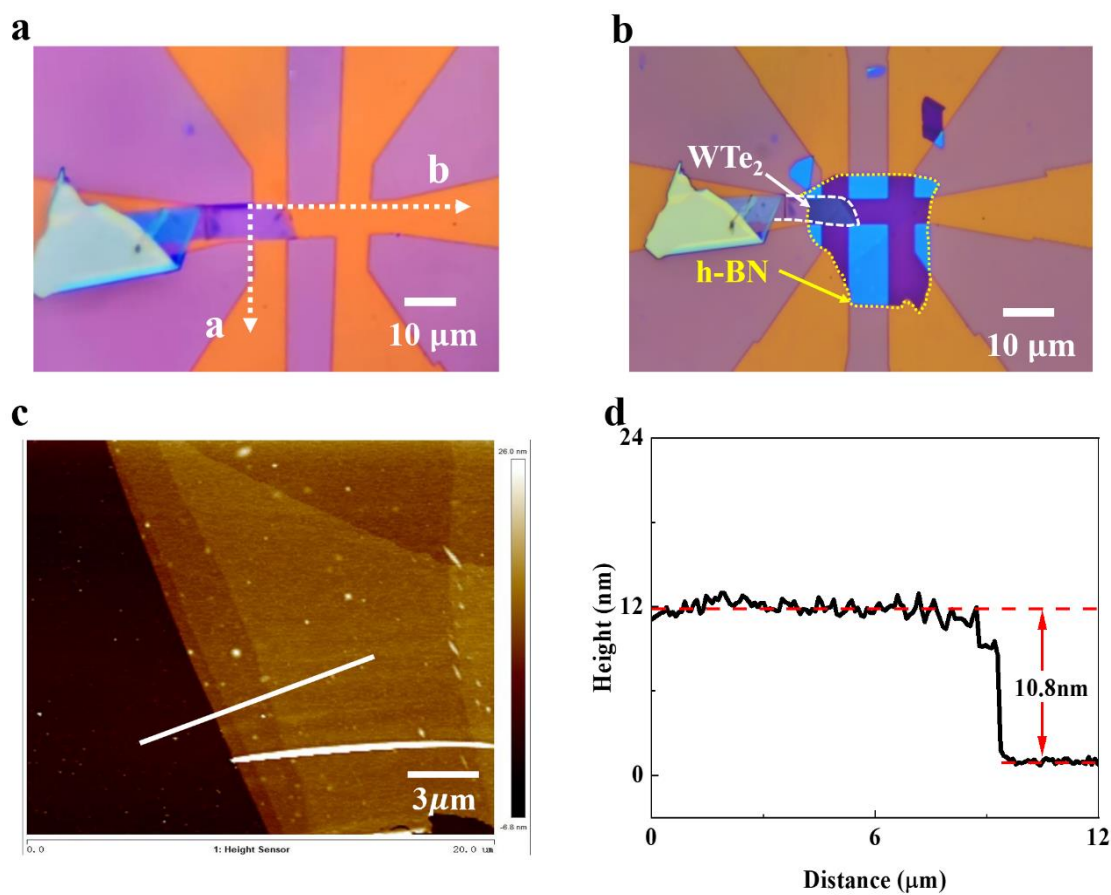

Figure S9. Optical picture of the device B without h-BN (a) and with h-BN (b). The dotted white and yellow boxes is the WTe$_2$ and h-BN flakes, respectively. Scale bar is 10 μm. (c) An AFM image of the WTe$_2$ flake with similar thickness as WTe$_2$ flake used for fabrication of device B. (d) A line cut [white line in (a)] from the edge of the WTe$_2$ flake, presenting an average thickness of about 10.8 nm.

**S10 AHR and Current-driven magnetization switching measurements of the device B.**

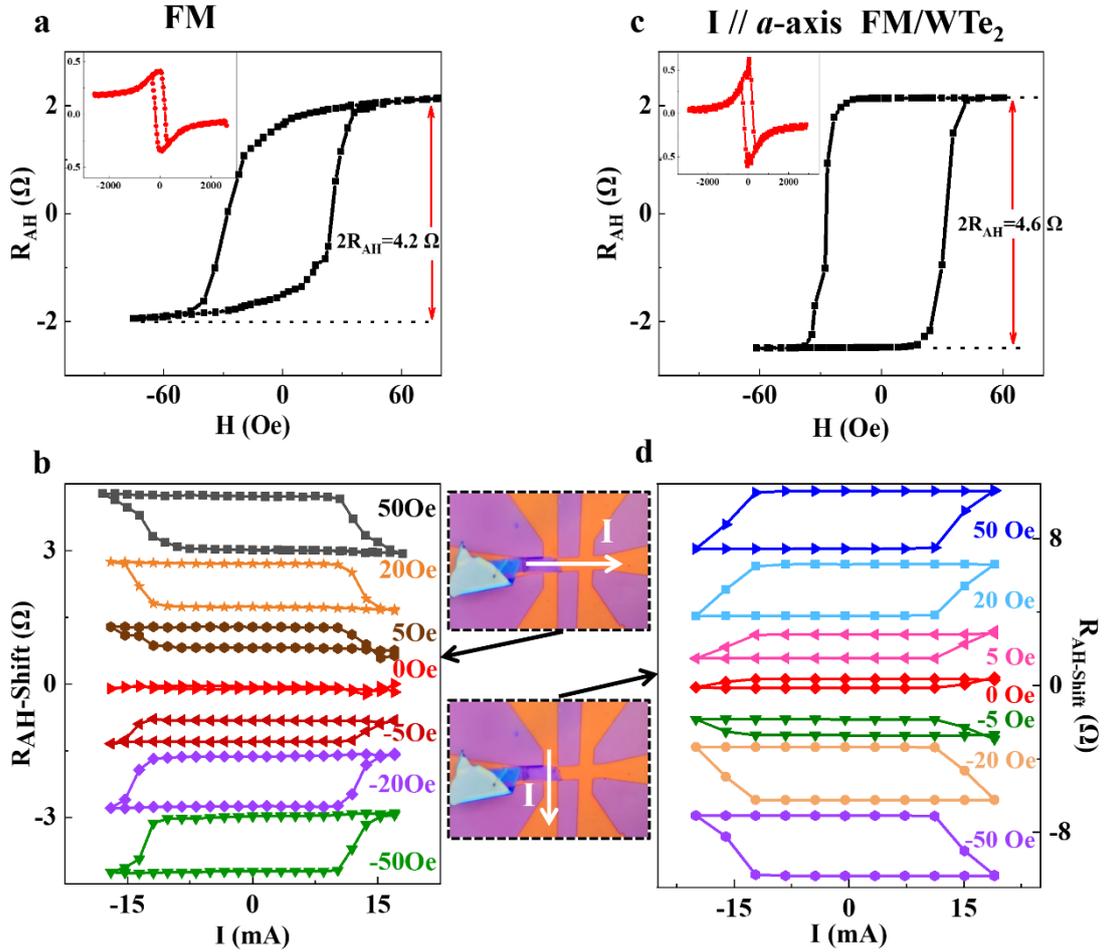

Figure S10. Current-driven magnetization switching measurements of the device B. The OP-AHR loop for the FM sample (a) and when current flows along the *a*-axis of the device B (c). The inset is the IP-AHR loop. $R_{AH}$-$I$ loops measured under various $H$ ranging from -50 to +50 Oe of the FM sample (b) and the device B (c) with the current along *a*-axis. (d) $R_{AH}$-$I$ loops under various H ranging from -50 to +50 Oe of device B when the current is applied to the *a*-axis.

**S11 Current-induced switching at zero magnetic field after the magnetization to up or down for the device B.**

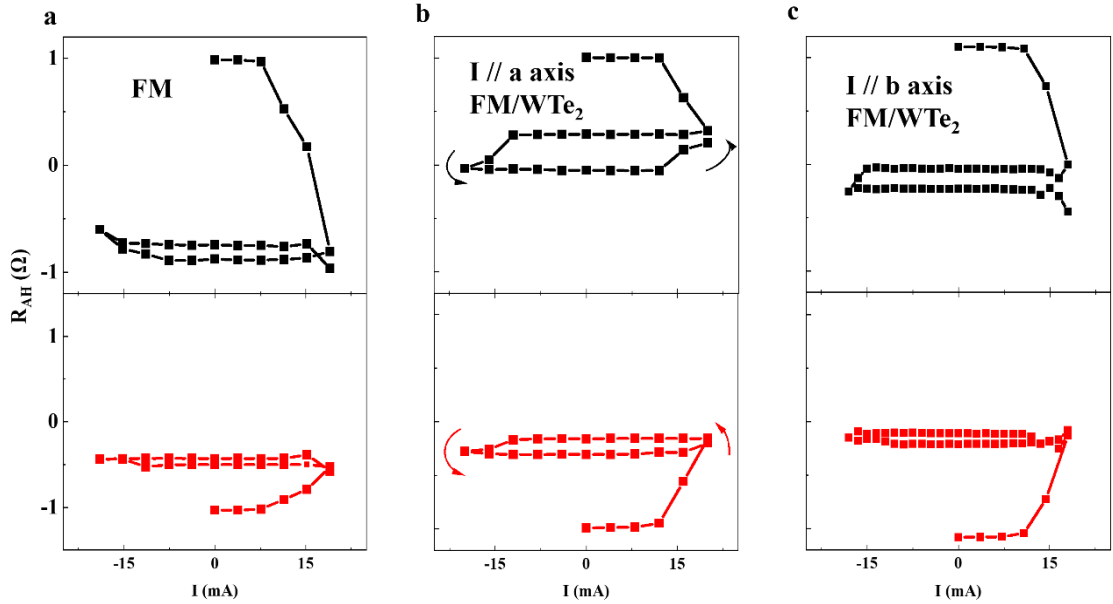

Figure S11. The initial state is magnetized up or down, current-induced switching curves for the FM sample (a), the device B with the current along the *a*-axis (b) and *b*-axis (c), respectively. When the current is applied to the *a*-axis (Fig. S11b), a counterclockwise loop is always observed, but not for the FM sample and device B with the current along the *b*-axis.

**Note S1. Calculation of current densities.**

Calculate the current density by the equation $J_0 = \frac{I}{S}$, where the $I$ is the current and $S$ is the area through which the current flows. For example, when I=3.5 mA, for FM layer, $S = 4.75$ nm $\times 5$ μm $= 2.375 \times 10^{-10}$ cm², so $J_0 = 1.47 \times 10^7$ A/cm². When the critical switching current $I_C$ is approximately 11.29 mA for the Pt/Co/Pt device at 50 Oe, which yields to a current density of $4.75 \times 10^7$ A/cm². Here, the mean current densities where $R_{AH}(J_c) = (R_{AH}^{max} + R_{AH}^{min})/2$ are defined as the critical switching current densities.

When the current is applied to *b*-axis, the resistance of the FM/WTe₂ bilayer is 332.80 Ω, and the resistance of the FM layer is 335.75 Ω. According to the parallel circuit model, calculated the resistance of the WTe₂ flake is 37877.15 Ω. When the current is 1 mA current along the *b*-axis, the current obtained by the FM and WTe₂ layers is 0.991 mA and 0.009 mA, respectively. The critical current density $J_C$ is equal to both that of the FM layer ($J_{FM}$) and the WTe₂ flake ($J_{WTe_2}$). When the critical current is 11.29 mA, we obtain the $I_{FM}$= 11.188 mA and $I_{WTe2}$=0.1016 mA. We then get that $J_{FM}$, $J_{WTe_2}$ and $J_{bc}$ ($J_{bc} = J_{FM} + J_{WTe_2}$) is equal to $4.711 \times 10^7$ A/cm², $(0.017 \pm 0.006) \times 10^7$ A/cm², and $(4.728 \pm 0.006) \times 10^7$ A/cm² with an in-plane magnetic field $H_x$=50 Oe, respectively. Note that, when the width to WTe₂ is equal to that of the FM layer, $J_{max}(WTe_2)$ =0.1016 mA/(8.7 nm×5 μm)=0.0233×10⁷ A/cm². and the minimum current density of WTe₂ flows through WTe₂ is the width of WTe₂. $J_{min}$=1mA/(8.7 nm×10.9 μm)=0.0107×10⁷ A/cm². Therefore, the confidence interval current density is $(0.017 \pm 0.006) \times 10^7$ A/cm². When the current is applied to *a*-axis of the device A, the resistance of *a*-axis at room temperature is about 1.8 times that of *b*- axis[2], so the resistance of *a*-axis of the WTe$_2$ flake is about 68178.87 Ω. When the current is 1 mA current along the *a*-axis, the current obtained by the FM and WTe$_2$ layers is 0.995 mA and 0.005 mA, respectively. When the critical current is 11.72 mA, we obtain the $I_{FM}$= 11.66 mA and $I_{WTe2}$=0.0586 mA. According to the similar method above, we can obtained $J_{FM}$, $J_{WTe_2}$ and $J_{ac}$ is equal to $4.91\times10^7$ A/cm$^2$, $(0.007\pm0.006)\times10^7$ A/cm$^2$ and $(4.917\pm0.006)\times10^7$ A/cm$^2$ with an in-plane magnetic field $H_x$=50 Oe, respectively.

**Note S2. Calculation of the $\chi_{SOT}$ and $\theta_{DL}$.**

The SOT fields per current density are obtained by using the formulas $\chi_{SOT}= H_{DL}/J$, where J is the current density[1]. We then calculated the damping-like effective Spin Hall angle ($\theta_{DL}$) using the formulas $\theta_{DL}= 2eM_s tH_{DL}/\hbar j = 2eM_s t\chi_{SOT}/\hbar$. where $e$ is the electron charge, $M_s$ is the saturation magnetization (158.4 emu/cm$^3$), $t$ is the FM layer thickness (t=3.75 nm) and $\hbar$ is the reduced Planck constant.

For only FM sample, the slope of the fitted curve in Fig. 4 (c) is 0.835Oe/mA. The current density $J$ is $4.211 \times 10^6$ A/cm$^2$, the $\chi_{SOT}$ (FM)=0.835/(4.211×10$^6$) Oe/(Acm$^{-2}$)=0.198×10$^{-6}$ Oe/(Acm$^{-2}$), thus the $\theta_{DL}(FM)$= 0.0036.

For the FM/WTe$_2$ sample, when the current is applied to *b*-axis, the slope of the fitted curve in Fig. 4 (f) is 0.821 Oe/mA, which contains two contributions respectively WTe$_2$ flake (denoted as $\chi_{SOT}(b-WTe_2)$) and FM layer. We consider the connection between the WTe$_2$ flake and FM layer to be parallel. When the current is 1 mA current along the *b*-axis of the device, the current obtained by the FM and WTe$_2$ layers is 0.991 mA and 0.009 mA, respectively. Thus, $\chi_{SOT}(b-WTe_2) = \frac{0.8214-0.835\times 0.991}{0.009\times S(WTe_2)}$ Oe/(Acm$^{-2}$), where $S(WTe_2)$ is the cross-sectional area value of the WTe$_2$ flake along the *b*-axis. Calculated $\chi_{SOT}(b-WTe_2)$ is equal to (-0.49±0.18)×10$^{-6}$ Oe/(Acm$^{-2}$). Therefore, $\theta_{DL}$ of the WTe$_2$ flake can be calculated as 0.0089±0.0033.

For the FM/WTe$_2$ sample, when the current is applied to *a*-axis, the slope of the fitted curve in Fig. 4(b) is 0.809 Oe/mA, which also contains two contributions respectively WTe$_2$ flake (denoted as $\chi_{SOT}(a-WTe_2)$) and FM layer. When the 1 mA current is applied to *a*-axis, the current obtained by WTe$_2$ is 0.005 mA and the current obtained by FM is 0.995mA. Therefore, $H_{DL}\chi H_{DL}\chi_{SOT}(a-WTe_2)$ and $\theta_{DL}$ of the WTe$_2$ flake can be calculated as $(3.04\pm1.13)\times10^{-6}$ Oe/(Acm$^{-2}$) and $0.055\pm0.020$.

**Table S1 Comparison of SOT switching efficiency and spin Hall angle.**

| Structure of Sample | Method of obtaining WTe$_2$ layer | Test method | Spin Hall angle | $\chi_{SOT}$ ×10$^{-6}$ Oe/(Acm$^{-2}$) | References |
|---|---|---|---|---|---|
| WTe$_2$/Pt/Co/Pt | Mechanical exfoliation | Second Harmonic | *a*-axis $\theta_{DL}$=-0.055±0.020 | 3.04±1.13 | This work |
| WTe$_2$/FeNi | Mechanical exfoliation | ST-FMR | $\theta_{SH}$ = 0.09 - 0.51 | 0.28±0.22 | [3] |
| WTe$_x$/Ti/CoFeB | Magnetron sputtering | Second Harmonic | $\theta_{DL}$ = -0.67±0.033 ×10$^5$ Ω$^{-1}$m$^{-1}$ | | [4] |
| WTe$_2$/Ta/CoTb | Mechanical exfoliation | Second Harmonic | $\theta_{DL}$= -0.14- -0.06 | 3.2±0.3 | [5] |